\documentclass[useAMS,usenatbib]{mn2e}

\usepackage{verbatim}
\usepackage{graphicx}
\usepackage{amssymb}
\usepackage{amsmath}
\usepackage[figuresright]{rotating}

\newcommand{\lam}{\Lambda}
\newcommand{\om}{\Omega_m} 
\newcommand{\apjl}{ApJL}
\newcommand{\hovm}{h{\rm Mpc}^{-1}}

\title[Power Spectra to 1\% Accuracy]{Power Spectra to 1\% Accuracy between 
Dynamical Dark Energy Cosmologies\thanks{Research undertaken as part of the Commonwealth Cosmology
Initiative (CCI: www.thecci.org), an international collaboration
supported by the Australian Research Council}} 
\author[Francis, Lewis \& Linder]{Matthew J. Francis$^{1}$\thanks{Email: mfrancis@physics.usyd.edu.au}, Geraint F. Lewis$^{1}$ and Eric V. Linder$^{2}$
\\
$^{1}$ School of Physics, University of Sydney, NSW 2006, Australia\\
$^{2}$ University of California, Berkeley Lab, Berkeley, CA 94720, USA
}

\begin{document}

\date{}

\pagerange{\pageref{firstpage}--\pageref{lastpage}} \pubyear{2007}

\maketitle

\label{firstpage}

\begin{abstract}
For  dynamical  dark  energy cosmologies we carry out a series of
N-body gravitational simulations, achieving percent level accuracy in
the relative mass power spectra at any redshift.  Such accuracy in
the power spectrum is necessary for next generation
cosmological mass probes.  Our matching procedure reproduces the CMB
distance to last scattering and delivers subpercent level power spectra
at $z=0$ and $z\approx3$.  We discuss the physical implications for
probing dark energy with surveys of large scale structure.
\end{abstract}

\begin{keywords}
methods:N-body simulations --- methods: numerical --- dark matter --- 
dark energy --- large-scale structure of Universe
\end{keywords}

\section{Introduction}

The  mass power  spectrum plays  a central  role in  understanding the
large  scale  structure  of  the  Universe and  the  cosmic  expansion
affecting the  development of  structure.  Cosmological probes  of the
growth  history or the  expansion history  such as  weak gravitational
lensing,  baryon acoustic  oscillations, or  galaxy/cluster abundances
depend crucially on accurate knowledge  of the mass power spectrum for
physical interpretation of the data from large scale surveys.

Most commonly the mass power spectrum is approximated by the `Halofit'
form of  \citet{Smith03}, and this  is used to  determine cosmological
parameters from  the data (or  estimate future precision  of parameter
extraction).  However, the  \citet{Smith03} formula is calibrated only
on  $\lam$CDM models,  and for  these has  a precision  of $\sim$10\%.
This will be insufficient for  future large structure surveys that aim
to explore the acceleration of the cosmic expansion and the properties
of dark  energy responsible for it.  \citet{huttak}  estimate that for
weak  gravitational  lensing surveys,  for  example,  1\% accuracy  in
knowledge of the mass power spectrum will be required.

N-body simulations provide a well-tested technique for calculating the
dark matter power spectrum at the percent level \citep{warren}.  While
this treats purely gravitational  forces, leaving out baryonic effects
including   heating  and   cooling,  this   should  be   a  sufficient
approximation     for      wavemodes     $k<3\,h\rm     Mpc^{-1}     $
\citep{jing,knox,whiteb}, or the scales larger than galaxies, that are
of  most   relevance  for   large  surveys.   However,   carrying  out
simulations  for  every   possible  cosmological  model  is  obviously
impractical.   If one could  devise a  mapping procedure  that matched
models  with the  same key  physical quantities,  ideally to  a single
class of cosmologies  like $\lam$CDM, then this would  greatly aid the
study of the cosmological  information carried by the distribution and
growth of large scale structure.

In Section \ref{background} we outline  the approach to such a mapping
procedure     and     compare     to    previous     work.     Section
\ref{Simulation-details}  describes  the  details of  the  simulations
performed   and   tests  carried   out.    The  qualitative   physical
consequences   of    the   mapping   are    interpreted   in   Section
\ref{Consequences}.  We  present the computational  results in Section
\ref{Results-Distance-Matched}   and   identify  several   interesting
features  in  the  wavemode  and  redshift  dependence  of  the  power
spectrum.  Physical  interpretation of these results  are discussed in
Section  \ref{Discussion},  with  conclusions  and  future  directions
summarised in Section \ref{Conclusion}.

\section{Dark Energy and Cosmic Structure}\label{background}

While  the influence of  dark energy  on the  linear growth  factor of
matter density perturbations can be calculated simply (see below), the
full, nonlinear mass power spectrum requires N-body simulations.  Such
simulation  studies of the  non-linear power  spectrum that  exist for
dark energy other than a cosmological constant have tended to be for a
constant dark energy  pressure to density ratio, or  equation of state
ratio                                                               $w$
\citep{1999ApJ...521L...1M,2004APh....22...19W,Linder&White,McDonald
et al}.

A few simulation studies have considered the effects of dynamical dark
energy on the power spectrum,  either through a parameterized time (or
scale factor  $a$) dependent  form $w(a)$ or  a specific  scalar field
potential \citep{klypin03,Maccio,2006astro.ph.10213M}.

Efforts  to  calibrate  non-linear  power  spectrum  fitting  formulas
include  early  work by  \citet{1999ApJ...521L...1M},  which was  much
improved upon with the  advances in computing power by \citet{McDonald
et al}. Both of these studies looked at modifying existing fits in the
case   of   constant   $w$   cosmologies.   \cite{Linder&White}   (LW)
investigated  the effects  of $w$  on the  non-linear  power spectrum,
searching  for  key  physical  quantities,  and  discovered  a  simple
matching  prescription  for  calculating  the  non-linear  mass  power
spectrum to within one to two percent.

This work  extends the study of  the dark energy effects  on the full,
non-linear mass  power spectrum to models with  dynamical dark energy,
utilizing  the model  independent, physically  motivated \citep{Linder
2003} evolving equation of state $w(a)=w_{0}+w_{a}(1-a)$.  At the same
time,  we employ  the approach  of seeking  central  physical matching
quantities that incorporate CMB  data through agreeing on the distance
to the last scattering surface.

In  studying the  non-linear power  spectrum of  mass  fluctuations, a
natural place to start is  with the linear power spectrum. The effects
of dark  energy on  the linear mass  power spectrum can  be calculated
through the relation
\begin{equation}
P(k,a)=\frac{D^{2}(a)}{D^{2}(a_{i})}P(k,a_{i})
\label{eq:powerfactor}
\end{equation} 
(see e.g. \citet{Coles & Lucchin}) with the growth factor $D(a)$ given 
by the formula (e.g. \citet{2002PhRvD..66h3515H,Linder&Jenkins})
\begin{equation}
D''+\frac{3}{2}\left[1-\frac{w(a)}{1+X(a)}\right]\frac{D'}{a}-\frac{3}{2}\frac{X(a)}{1+X(a)}\frac{D}{a^{2}}=0\label{eq:growth factor} 
\end{equation} 
with derivatives with respect to scale factor $a$, and where $X(a)$ is
the  ratio  of   the  matter  to  dark  energy   densities,  given  by
$X(a)=\frac{\Omega_{m}}{1-\Omega_{m}}e^{-3\int_{a}^{1}d\ln   a'w(a')}$,
with $\om$ the dimensionless present matter density.

The non-linear power  spectrum cannot be written in  terms of a simple
differential  equation and  requires  the use  of  large volume,  high
resolution,  N-body simulations.  These are  computationally expensive
and  therefore accurate  semi-analytic fitting  formulas  derived from
simulation  results  are a  valuable  tool.  The  most widely  adopted
current   formula,  sometimes   called  Halofit,   was   presented  in
\citet{Smith03}.   This formula  is  motivated by  the  halo model  of
structure growth with  free parameters in the function  set by fitting
to a large suite of simulations.  All these simulations, however, were
of cosmological constant,  $w=-1$, cosmologies. \citet{McDonald et al}
produced a  fitting formula as  a multipolynomial series  for constant
$w$ models,  intended to  be used  to modify the  Smith et  al result.
This modification was estimated to be accurate to within a few percent
in the range of cosmologies encompassed by the simulation grid.

Taking  a  different approach,  LW  demonstrated  that  when the  {\it
linear\/}  growth  factors  between  different $w=$const  models  were
matched at a high redshift point  as well as at $z=0$, by compensating
with  other  cosmological  parameters,  the {\it  non-linear\/}  power
spectrum from  N-body simulations also  matched to much better  than a
percent at those redshifts, as well  as matching to one to two percent
at any  intermediate redshift.  Additionally,  LW also found  that the
distance  to the surface  of last  scattering, $d_{\rm  lss}$, closely
matched   when  their  growth   matching  criteria   was  implemented,
preserving CMB  constraints.  With  this formalism the  power spectrum
for  a  dark energy  model  can be  matched  to,  say, a  $\Lambda$CDM
cosmology.  Hence one can  either use an appropriately matched Halofit
result  or  carry  out  a  vastly  reduced  suite  of  only  $\lam$CDM
simulations  to  achieve  the  desired  accuracy  on  the  mass  power
spectrum.

This  article  concentrates on  developing  accurate  matching of  the
non-linear mass  power spectrum for dynamical dark  energy models.  We
employ  a somewhat  different matching  procedure from  LW, explicitly
matching the  distance to  CMB last scattering  $d_{\rm lss}$  and the
mass fluctuation amplitude $\sigma_8$  at the present and studying the
effect on  the growth.  In  this respect, our approach  is essentially
the converse of the LW approach.  The geometric factor of the distance
to CMB  last scattering suffices to incorporate  substantially the CMB
constraints on  the dark energy  parameters.  Since dark energy  had a
negligible  density in the  early universe  (except in  special, early
dark energy  models, e.g.\ see  \citet{wettearly,doran,linearly}), the
physical size and nature of features in the CMB at the surface of last
scattering  is   largely  insensitive   to  the  properties   of  dark
energy. However, the angular size  of such features is set through the
angular diameter distance, which does depend on the properties of dark
energy,  since it  relates to  the expansion  history of  the universe
$a(t)$. Therefore, dark energy models  giving the same distance to the
last scattering surface are largely degenerate with respect to the CMB
(some differences  relating to secondary anisotropies such  as the ISW
effect  remain, see  \cite{CMBan}). For  a given  dynamic  dark energy
model $(w_0,w_a)$, there is a corresponding constant equation of state
$w_{\rm eff}$, say, that gives the same $d_{\rm lss}$ as the dynamical
model, holding all other cosmological parameters (such as the physical
matter density  $\om h^2$) fixed.  This article  examines the relation
between the  non-linear mass  power spectra of  the dynamical  and the
$w_{\rm eff}$ models.  Once a tight correspondence is established, one
can then  either employ  a constant $w$  fitting formula such  as from
\citet{McDonald  et  al}, carry  out  only  a  suite of  constant  $w$
simulations, or adjust the other cosmological parameters such that one
chooses $w_{\rm eff}=-1$ and  requires only $\lam$CDM simulations.  We
discuss these alternatives further in \S\ref{Conclusion}.

\section{Simulation details}\label{Simulation-details}

The  simulations  were  performed   using  the  GADGET-2  N-body  code
\citep{Gadget code}, modified  to incorporate the background evolution
$a(t)$  appropriate   for  dynamical  dark   energy  cosmologies  with
$w(a)=w_{0}+(1-a)w_{a}$.  Fiducial simulation  runs use $256^{3}$ dark
matter particles  in a $256\,h^{-1}$Mpc periodic box  with a $512^{3}$
force grid;  the initial redshift  was $z=24$ and the  force softening
was set to a constant  co-moving length of $60\,h^{-1}$Kpc. In order to
check   numerical   convergence,  runs   were   also  performed   with
combinations of box size and particle number a factor of 2 greater and
smaller than the fiducial. In addition, runs checking convergence were
performed for numerical parameters including the start time, softening
length,  PM grid  spacing, time  and  force accuracy  and tree  update
frequency. The ratio of power between the different dark energy models
were  largely insensitive  to these  parameters, changing  by  a small
fraction of a percent out to $k < 3\,h {\rm Mpc}^{-1}$.

The linear matter power spectra  used to create the initial conditions
were calculated using CAMB \citep{CAMB code paper}. Initial conditions
were  generated from  the  power  spectrum using  part  of the  GRAFIC
program within the COSMICS package \citep{COSMICS code paper}.

For each set of distance-matched  runs, the same input power spectrum,
generated by  CAMB using  the $w={\rm constant}$  model, was  used for
each model.   In order to match  the amplitude of  linear growth today
(identical   $\sigma_8$  at  $z=0$)   for  simulations   of  different
cosmologies,  the initial  density and  velocity perturbations  of the
particles were scaled in the Zel'dovich approximation using the linear
growth  factor  ratio  $D(a_{\rm  start})/D(a=1)$  for  the  different
models.  This ansatz  for initial conditions is robust  as long as the
dark energy does not change the  shape of the linear power spectrum at
$z_{\rm start}$, i.e.\  the dark energy plays little  role in the very
early universe. We have verified this to high accuracy using a version
of  CAMB   modified  for   $(w_0,w_a)$  models  without   dark  energy
perturbations.    Note   that  in   the   presence   of  dark   energy
perturbations,  the   initial  power   spectrum  over  our   range  of
$k=0.1-3\,\hovm$ is affected by less  than 1\% for constant $w$ models
that   are   {\it   not\/}   distance   matched.    We   expect   that
distance-matched $(w_0,w_a)$ models  with perturbations will show less
effect but future work will address this.

The calculation of the power  spectrum in simulations outputs used the
`chaining the power' method described in \citet{Smith03} utilising the
cloud in  a cell  assignment scheme. No  correction was made  for shot
noise, as the quantity of interest  was the ratio of the power between
different  models.   See  \citet{McDonald  et  al}   for  an  extended
discussion  of  the usefulness  of  taking  power  spectrum ratios  to
eliminate  many numerical  errors  in  this type  of  study. See  also
\citet{2005APh....24..334W}.

All  simulations  in  this   paper  used  the  best  fit  cosmological
parameters  from  \citet{Spergel}  of  $\Omega_m =  0.234$,  $h=0.74$,
$\Omega_b=0.0407$  and   $\sigma_8  =   0.76$  in  a   flat  $\lam$CDM
universe. For  each set of  simulations, a constant equation  of state
$w_{\rm eff}$ is selected and  several values for the parameters $w_0$
and $w_a$ that maintained the  same $d_{\rm lss}$ were calculated. One
consequence of this methodology is  that these $w(a)$ models cross the
value  $w=-1$  at  some   point  in  cosmic  history.   Debate  exists
surrounding  the physical  validity  of crossing  between the  phantom
regime,  defined  as $w<-1$,  and  $w>-1$.   This  will eventually  be
settled by a microphysical theory  for dark energy, rather than merely
a phenomenological description.  With this  in mind we do not consider
the issue of phantom energy and phantom crossing further.

We select three values, $w_{\rm eff}=-0.9,-1,-1.1$, as the foundations
for our  comparison of  $w(a)$ cosmologies.  This  range is  in accord
with constraints  on constant $w$ from current  cosmological data sets
\citep{Spergel,seljak} and  provides a reasonable  variety for testing
the matching procedure.  For each constant $w$ model, simulations were
carried out for four more  $w(a)$ models with matching distance to the
LSS. The  dark energy  models used are  summarized in  Table \ref{dist
matched table}.

\begin{table}
\begin{center}
\begin{tabular}{|c|c|c|c|c|c|}
\hline
\multicolumn{2}{|c|}{$w_{\rm eff}=-0.9$} \qquad &
\multicolumn{2}{|c|}{$w_{\rm eff}=-1.0$} \qquad &
\multicolumn{2}{|c|}{$w_{\rm eff}=-1.1$} \qquad
\\
\hline
$w_0$&
$w_a$&
$w_0$&
$w_a$&
$w_0$&
$w_a$
\\
\hline
\hline
-1.1   &
0.620  & 
-1.2   &
0.663  &
-1.3   &
0.707 
\\
\hline 
-1.0   &
0.319  &
-1.1   &
0.341  &
-1.2   &
0.363 
\\
\hline
-0.8   &
-0.336 &
-0.9   &
-0.359 &
-1.0   &
-0.381
\\
\hline
-0.7   &
-0.686 &
-0.8   &
-0.732 &
-0.9   &
-0.778 
\\
\hline
\end{tabular}
\caption{Distance  Matched Models.  Simulations were  carried  out for
five models  (including $w=w_{\rm eff}$)  for each of three  values of
$w_{\rm  eff}$,  where all  five  models  in  a column  had  identical
distances to CMB last scattering.}\label{dist matched table}
\end{center}
\end{table}

\section{The Consequences of Distance Matching} \label{Consequences}

\begin{figure}
\includegraphics[
  scale=0.35,
  angle=-90]{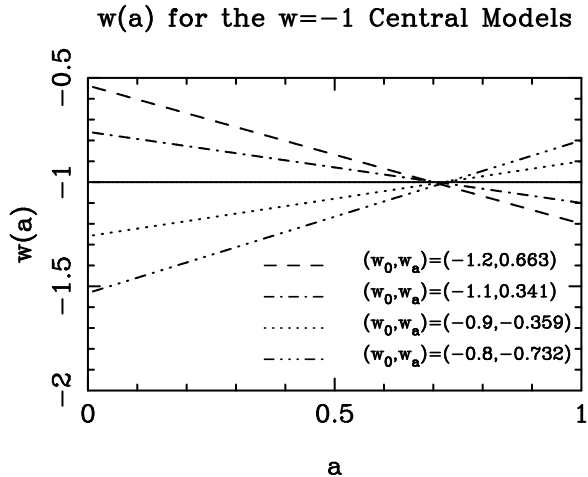}
  \caption{Dark  energy equation  of  state vs.\  $a$  for the  models
matched   to   the   CMB   last  scattering   surface   distance   for
$w=-1$.}\label{wfig}
\end{figure}

If the simple distance matching procedure outlined in this paper is to
succeed in  producing a  good match in  the matter power  spectrum, it
might  be expected  to keep  a  range of  physical conditions  similar
through cosmic history. This is indeed what is found. In particular, a
variety  of physical  quantities  exhibit pivot  or crossover  points,
indicating not only near equality at that epoch, but a tendency toward
agreement of quantities integrated over cosmic history.  For instance,
Figure  \ref{wfig} plots  $w(a)$  for the  four  models with  matching
distance to the $w=-1$ model;  there is a clear epoch at $a\approx0.7$
where the values all cross $w=-1$.

\begin{figure}
\includegraphics[
  scale=0.35,
  angle=-90]{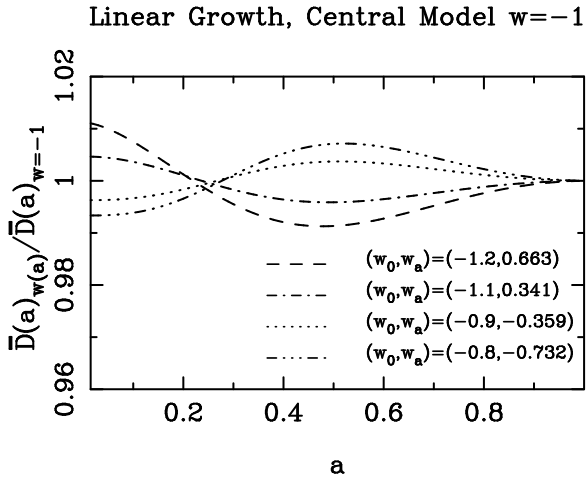}
\caption{Ratio of the linear growth factor $\bar D(a) \equiv D(a/D(1)$
relative  to the  central  $w=-1$  model for  the  $w(a)$ models  with
matched $d_{\rm lss}$.}\label{wm1linplot}
\end{figure}

The linear growth factors of  the various distance matched models also
closely track each other.  While the linear growth is matched at $a=1$
by construction, there  is an additional epoch at  high redshift where
the linear  growth of all  associated models closely matches.  For the
$w_{\rm eff}=-1$  set of models shown in  Figure \ref{wm1linplot}, the
matching point is  $a=0.24$ or $z=3.12$. The other  two sets of models
match at  a similar  value and show  a similar trend.  This behaviour
illustrates the converse of what  was found in LW, where $d_{\rm lss}$
was found to closely match when  the linear growth was matched at some
high redshift point. The crossover is important, since rather than all
models diverging  from the  $z=0$ match, the  curves track  each other
relatively closely.  Since the non-linear power spectrum is intimately
tied to the linear behavior, this  provides hope that a mapping of the
full power spectrum between models can be realized.

\begin{figure}
\includegraphics[
  scale=0.35,
  angle=-90]{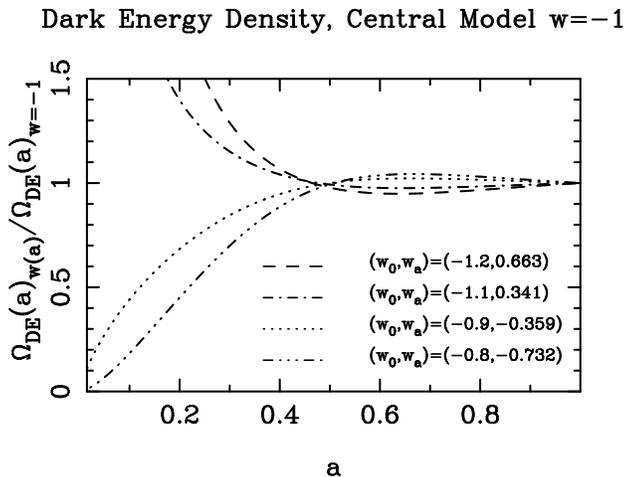}
\caption{Ratio of $\Omega_{\rm DE}(a)$  relative to the central $w=-1$
model    for    the     $w(a)$    models    with    matched    $d_{\rm
lss}$.}\label{wm1xplot}
\end{figure}

The growth  crossover behaviour results from  a change in  sign of the
relative rate  of growth  at $a\approx0.5$. In  the linear  regime the
growth of  fluctuations can  be seen as  a balance between  the mutual
gravitational  attraction of  the overdensity,  which is  amplified by
higher mean  matter density,  and the expansion  rate of  the universe
characterised by $H(a)$, which acts like a frictional term, suppressing
growth  the  higher  the  expansion  rate. In  this  picture,  greater
relative matter domination  at a particular epoch will  produce a more
rapid growth  rate at  that time compared  to a less  matter dominated
model. With this  in mind it is worth  comparing the matter domination
history of  our suite of cosmologies. Figure  \ref{wm1xplot} shows the
relative dark  energy density  $\Omega_{de}(a)$ for the  $w(a)$ models
matched to $w=-1$.

Comparing  figures  \ref{wm1linplot} and  \ref{wm1xplot}  we see  that
where the relative dark energy density is higher, the relative rate of
linear growth  is lower.   For instance, the  long dashed model  has a
more negative  slope in the region $a<0.5$  in figure \ref{wm1linplot}
than the other models, corresponding  to a higher relative dark energy
density  in  this  region  as  shown in  figure  \ref{wm1xplot}.   The
magnitude of  growth is greater initially  in the long  dashed model in
order to achieve the match at $z=0$, however the growth initially grows
more slowly in comparison to the  other models.  The change in sign of
the relative growth rates  at $a\approx0.5$ in figure \ref{wm1linplot}
corresponds to  the crossover point  in relative matter  domination in
Figure~\ref{wm1xplot}.   The energy density,  like the  linear growth,
exhibits a striking crossover point, again keeping physical conditions
similar between models throughout cosmic history.

In a chain of related  conditions, the crossover point of the equation
of state $w(a)$ (see  Figure~\ref{wfig}) causes the convergence of the
dark energy  density $\Omega_{de}(a)$ (see  Figure~\ref{wm1xplot}, and
the  crossover of  $\Omega_{de}(a)$ leads  to the  convergence  of the
growth  $D(a)$  (see Figure~\ref{wm1linplot}),  which  then creates  a
crossover in  the growth at higher  redshift.  This in  turn will keep
the non-linear power spectrum  closely matched between the models over
the entire range $z=0-3$.

\section{Results}\label{Results-Distance-Matched}

\begin{figure}
\includegraphics[
  scale=0.35,
  angle=-90]{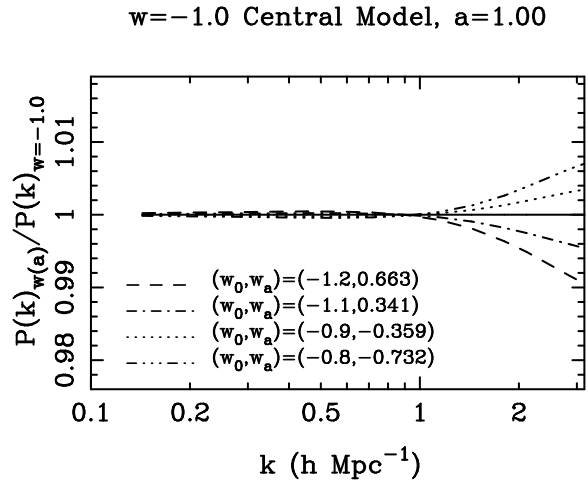}
\caption{Ratio of the non-linear mass power spectrum at z=0  relative 
to the $w=-1$ model for models with matched $d_{\rm lss}$.}\label{wm1power}
\end{figure}

\begin{figure}
\includegraphics[
  scale=0.35,
  angle=-90]{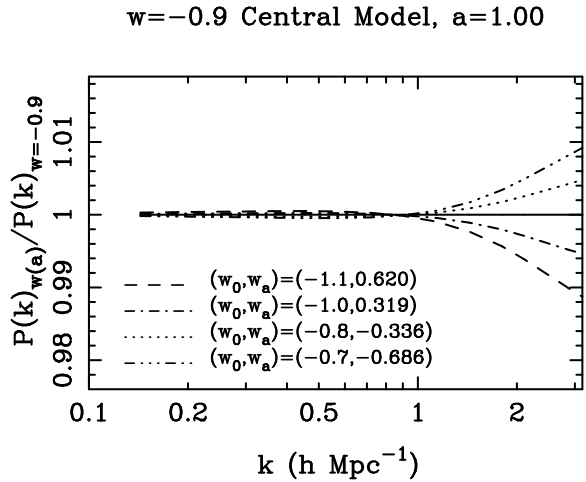}
\caption{As Fig.~\ref{wm1power}, for the central model 
$w=-0.9$.}\label{wm0.9power}
\end{figure}

\begin{figure}
\includegraphics[
  scale=0.35,
  angle=-90]{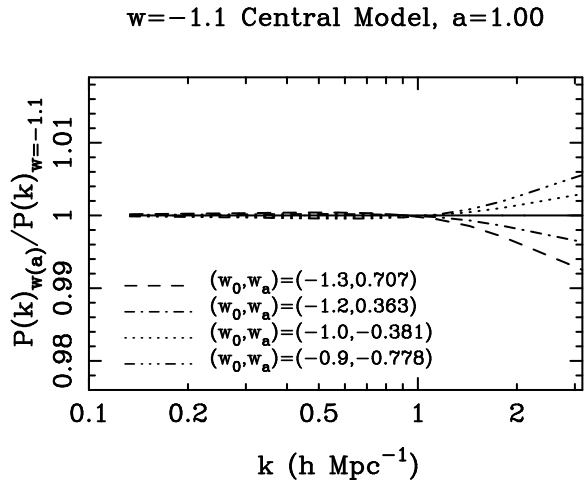}
\caption{As Fig.~\ref{wm1power}, for the central model 
$w=-1.1$.}\label{wm1.1power} 
\end{figure}

The  ratio  of power  measured  in  the  simulation outputs  at  $z=0$
relative to  the central $w_{\rm eff}$  model for each of  the sets of
simulations is  given in Figures  \ref{wm1power}, \ref{wm0.9power} and
\ref{wm1.1power}.   The  most  outstanding  result  is  the  excellent
agreement between  the mapped  power spectra, at  the 0.1\%  level for
$k<1\,\hovm$  and $\lesssim1\%$  for $k<3\,\hovm$  ($\lesssim0.5\%$ at
the  higher $k$  for the  less  rapidly varying  dark energy  models).
These figures  show very similar  trends, regardless of  the fiducial
$w_{\rm  eff}$  model  chosen.  Since  the trends  are  similar,  the
remaining figures will  show the results for the  $w=-1$ central models
only, for brevity.

The  simulations shown  in  these figures  started  with an  identical
realisation of  initial conditions, albeit scaled with  respect to the
linear growth factor to produce matched linear growth at $z=0$. As the
deviations between models  are small, it is important  to take care to
ensure that  any features are real  and not the results  of a spurious
numerical    effect.     The    convergence   tests    described    in
Section~\ref{Simulation-details}  addressed the  effects  of numerical
parameters
\footnote{Changing particle  resolution did cause  a slight systematic
shift in features seen in the  power spectrum ratios. The onset of the
dispersion     between     the      models     seen     in     figures
\ref{wm1power}-\ref{wm1.1power}  at $k\sim 1  \hovm$ shifted  to lower
$k$  with   reduced  particle  resolution  and  higher   $k$  with  an
increase. This shift was of order  $0.1$ in log $k$ for factors of two
differences in particle resolution.   We cannot fully account for this
numerical effect, however the difference  in power at a given $k$-mode
due to the shift is at  most $\sim 0.1\%$ and since subpercent effects
are beyond the ability of $N$-body simulations to probe accurately, we
do not believe this effect is of significant consequence.}.

\begin{figure}
\includegraphics[
  scale=0.35,
  angle=-90]{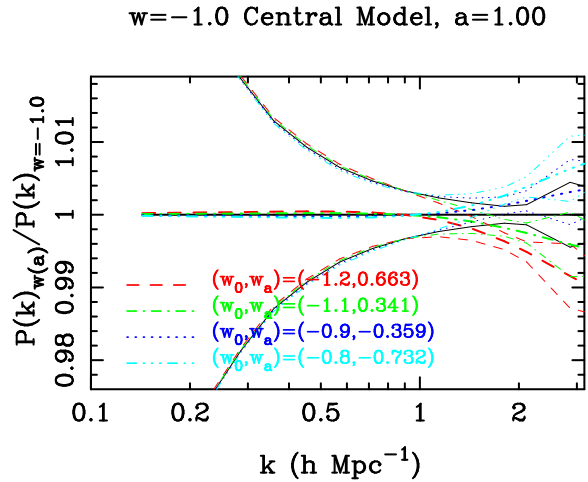}
\caption{As Fig.~\ref{wm1power}, with rms sampling errors included, 
shown by thin lines of the same line style as each model.}\label{errplot}
\end{figure}

There  are also  two other  potential  sources of  error, the  limited
volume  of  the  simulation  box  and  the  error  sampling  error  in
calculating the power spectrum of the simulation snapshots.  Care must
be take that  these effects are not causing  spurious results.  Figure
\ref{errplot}  shows   the  results  for  the   $w=-1$  central  model
simulations  with rms  sampling  errors from  the  FFT power  spectrum
calculation  plotted. For  clarity these  have been  omitted  from the
other plots, however the errors are similar in all cases.

\begin{figure}
\includegraphics[
  scale=0.35,
  angle=-90]{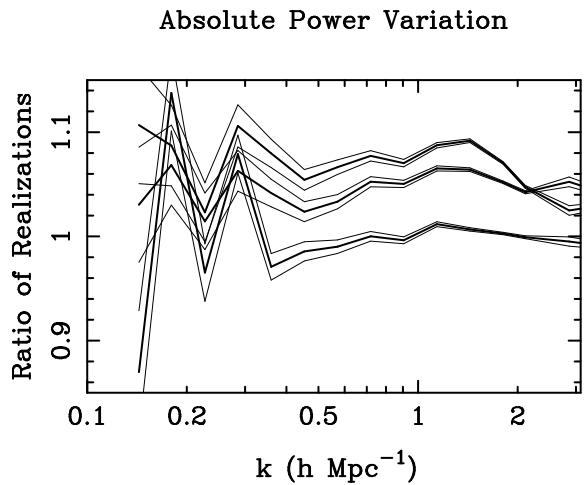}
\caption{The   effect  on   the  power   spectrum  due   to  different
realisations.   Displayed are  the power  in three  realisations  of a
single cosmological model, $w=-1$, as a ratio to the power in a fourth
realisation. The  rms errors for  each power spectrum  calculation are
also  shown. As  expected the  finite  volume error  decreases as  $k$
increases  due  to the  greater  number  of  modes present  at  higher
wavenumbers.}\label{real_scatter}
\end{figure}

From this  plot the  deviation between models  is roughly a  factor of
$1-2$ that of the rms error.  In order to verify that the effects seen
are genuinely  due to  the difference in  dark energy  models, another
three  sets of  simulations  with the  same  parameters but  different
realisations  were  performed. The  scatter  in  the calculated  power
spectrum   due  to   different   realisations  is   shown  in   Figure
\ref{real_scatter}.  The scatter in  this figure  clearly demonstrates
the inability to accurately determine  the absolute power with the box
sizes  and number of  realisations used  in this  study due  to finite
volume errors.

\begin{figure}
\includegraphics[
  scale=0.35,
  angle=-90]{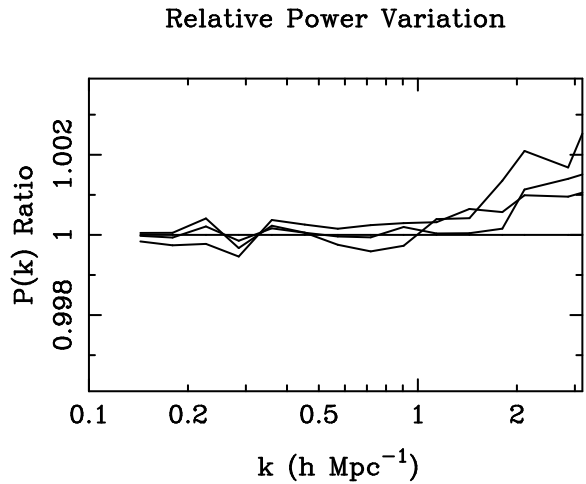}
\caption{The  effect  of different  realisations  on  the  ratio of  a
dynamical  dark  energy power  spectrum  to  the associated  $w=const$
model.  Shown  are 4  realisations of the  ratio between  the strongly
time varying  $(w_0,w_a)=(-1.2,0.663)$ model  and the $w=-1$  model at
$z=1$,  plotted  as  a ratio  of  the  main  realisation used  in  the
paper. This is a typical example of the magnitude of the variation due
to different realisations.}\label{rel_scat}
\end{figure}

From  the  figures shown  it  is  clear  that the  difference  between
realisations  for a  single  cosmological model  is  greater than  the
difference within a single realisation for the different models.  This
makes accurate modeling of the absolute value of the power spectrum an
extremely  challenging   task,  requiring  larger   boxes,  many  more
realisations and highly detailed consideration of sources of numerical
error.  Instead, we are interested in the effect of dark energy models
relative to  one another and therefore  what is important  is how much
the   {\it    ratios\/}   between    models   (such   as    shown   in
Figures~\ref{wm1power}-\ref{wm1.1power})  are  affected  by  different
realisations.  Fortunately  the effect  of  different realisations  is
vanishingly  small as  seen  in Figure~\ref{rel_scat},  which shows  a
typical  example of  the variation  in  power ratios  across the  four
realisations used.

From  Figure~\ref{rel_scat}  we can  be  confident  that the  computed
non-linear power  spectra ratios are not visibly  affected by spurious
finite  volume errors or  effects due  to the  FFT calculation  of the
power spectrum.

For cosmological structure  probes, we are interested not  just in how
well  we  can predict  the  power spectrum  at  $z=0$  but across  all
redshifts. In the  simulations performed, data was output  at a number
of times.  From Figure~\ref{wm1linplot},  two epochs are of particular
interest. The first is the  crossover in linear growth at $a=0.24$ and
the  second is  at  $a=0.5$ where  the  linear growth  is most  varied
between models.  The latter corresponds  to $z=1$, which  is extremely
relevant to  a number of  forthcoming cosmological surveys.   Hence an
accurate estimation  of power here, provided by  the distance matching
scheme, is of great  importance for understanding possible constraints
on dark energy cosmologies.

\begin{figure}
\includegraphics[
  scale=0.35,
  angle=-90]{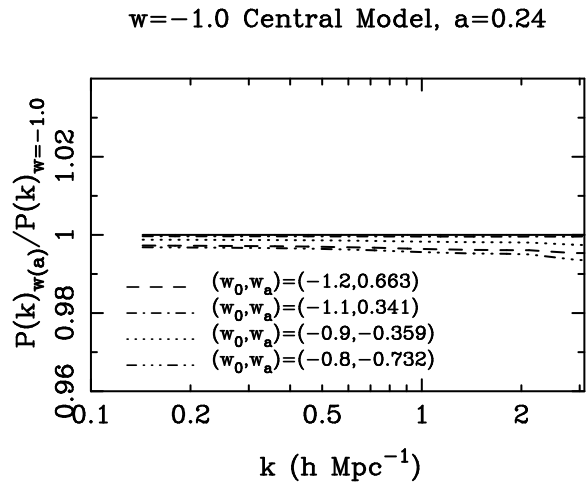}
\caption{As   Fig.~\ref{wm1power},   but   at  $a=0.24$   where   from
Fig.~\ref{wm1linplot} the linear growths closely match (slightly lower
than for $w=-1$).}\label{power0}
\end{figure}

\begin{figure}
\includegraphics[
  scale=0.35,
  angle=-90]{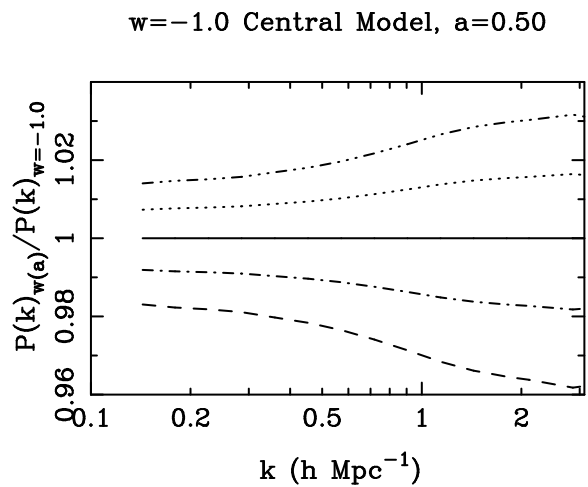}
\caption{As Fig.~\ref{wm1power}, but at $a=0.5$ where from 
Fig.~\ref{wm1linplot} the linear growths are most divergent.  
Much of the difference in power comes from this difference in 
linear power.}\label{power3}
\end{figure}

The ratio of power measured in the simulations boxes at these epochs 
are shown in Figures~\ref{power0} and \ref{power3}. 

\begin{figure}
\includegraphics[
  scale=0.35,
  angle=-90]{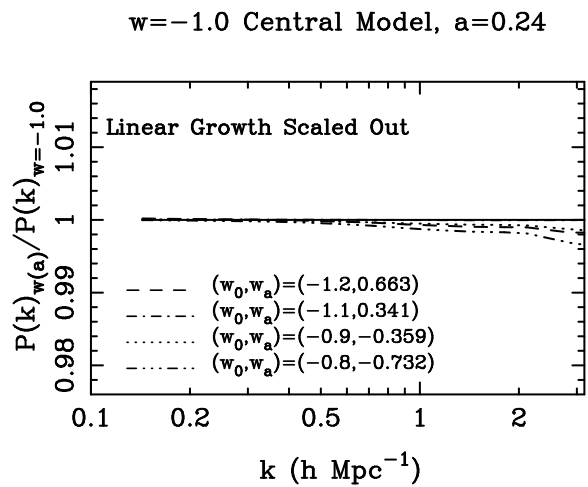}
\caption{As Fig.~\ref{power0}, but with the linear growth difference 
scaled out.  Note the reduced y-axis scale relative to 
Fig.~\ref{power0}.}\label{spower0} 
\end{figure}

\begin{figure}
\includegraphics[
  scale=0.35,
  angle=-90]{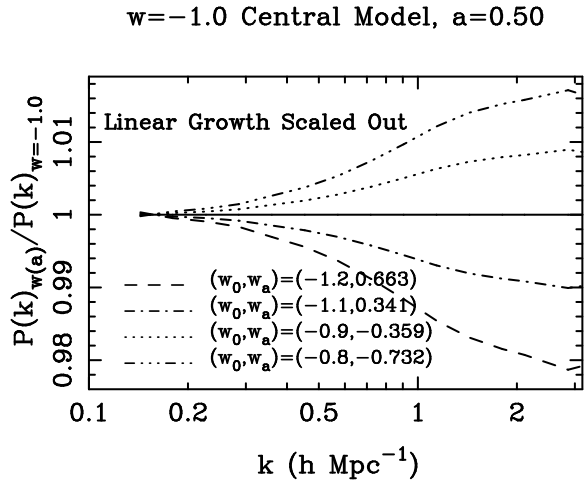}
\caption{As Fig.~\ref{power3}, but with the linear growth difference 
scaled out.  Note the reduced y-axis scale relative to 
Fig.~\ref{power3}.}\label{spower3} 
\end{figure}

We  can  do  even better,  however,  by  realizing  that much  of  the
difference in  power, particularly at $a\approx0.5$,  can be accounted
for  by  the  difference  in  linear  power.   Scaling  this  out  via
Equation~\ref{eq:powerfactor},   the   results   are   as   shown   in
Figures~\ref{spower0} and \ref{spower3}.

From these figures we can see that the combined distance and growth 
matching procedure is generally accurate to better than 1\%.  The 
greatest deviation found in all simulation outputs is 2\% for 
$k\approx3\,h{\rm Mpc}^{-1}$ at $a=0.5$.

\begin{figure}
\includegraphics[
  scale=0.35,
  angle=-90]{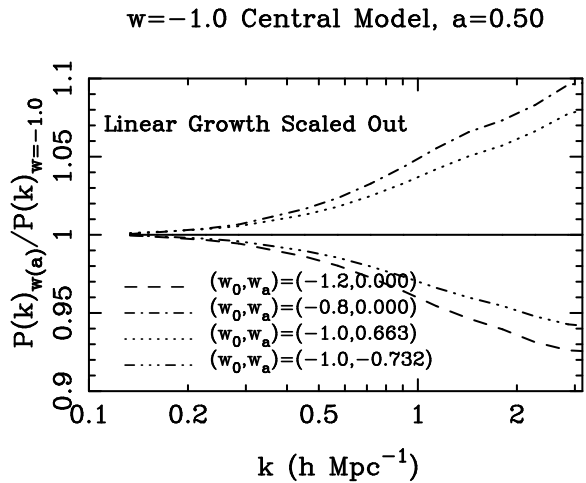}
\caption{The  ratio of  the non-linear  power in  several non-distance
  matched cosmologies to a  $\Lambda$CDM model at $z=1$. The amplitude
  of linear growth at $z=0$ is  the same for all models as in previous
  figures. Note that the  divergence between models increases with $k$
  but that the  divergence begins at a lower  $k$ and is significantly
  greater than with the distance matched models shown previously}
 \label{nonmatch} 
\end{figure}

Since the  results shown  thus far display  a good match  for distance
matched  models, it  is worth  considering how  much  improvement this
matching achieves  compared to arbitrary  dark energy models  that are
not distance matched. In other words, how much of a role do the values
of  the dark  energy parameters  play in  structure formation,  if all
other parameters are kept fixed? Figure \ref{nonmatch} shows the ratio
of  power   at  $z=1$  between   a  $\Lambda$CDM  model   and  several
non-distance matched  models with the same linear  growth amplitude at
$z=0$. The  divergence between  these models is  significantly greater
than what is  seen in the distance matched models  (and of course they
will disagree  on the CMB),  illustrating the improvement  achieved by
this simple scheme.

\section{Evolving Dark Energy and Structure Growth}\label{Discussion}

The  matching prescription  used in  this article  produces  a mapping
between the  matter power spectra of dark  energy cosmologies accurate
to $\lesssim1\%$  over a wide range  of wavemodes and a  wide range of
cosmic  history.  This  agreement  1) indicates  that simple  physical
quantities  determine the  nonlinear power  spectrum over  this range,
leading  to the  prospect of  understanding structure  formation  on a
fundamental level even in dynamical dark energy cosmologies, 2) points
the  way  to  significant  advances  in  computational  efficiency  by
reducing the dimension of the grid of simulations necessary to produce
accurate estimations  of power  spectra required for  interpretation of
cosmological  probes  such   as  weak  gravitational  lensing,  baryon
acoustic oscillation, and other  large scale structure surveys, and 3)
identifies a degeneracy that makes it difficult to distinguish between
models lying on a  particular subsurface of the cosmological parameter
space.

To try to ameliorate the degeneracy, we note that an evolving equation
of state does imprint a  small but systematic effect on the non-linear
matter    power    spectrum.    The    general    trend    shown    by
Figures~\ref{wm1power}, \ref{wm0.9power}  and \ref{wm1.1power} is that
dark energy with  a less negative value today  but more negative value
at high redshift (i.e.\ negative $w_a$) gives greater non-linear power
at $k\gtrsim  1\,\hovm$ than  its $d_{\rm lss}$-matched  $w_{\rm eff}$
model.   Similarly,  more  negative  equations  of  state  today  with
positive $w_a$ possess less power in the same range. This deviation is
however   relatively  small,   remaining  less   than  $2\%$   out  to
$k=3\,\hovm$.  Even  so, this partial  degeneracy is not  too worrying
since  it can  readily be  broken by  other  cosmological dependencies
(e.g.\ the geometric distance  dependence entering with the mass power
spectrum  into the  weak lensing  shear power  spectrum or  the baryon
acoustic  scale) or  by complementary  cosmological probes.   Thus the
model mapping technique does not  appear to have any real drawbacks to
detract from its physical and computational advantages.

Elaborating on the physical import  of the mapping, a striking feature
is the marked difference at $z=0$ between $k<1\,\hovm$ where the power
spectra between models match  extremely closely and $k>1\,\hovm$ where
they diverge.  This seems to  suggest a transition between  the linear
region at low $k$ where by design the power should be matched, and the
fully  non-linear  region at  high  $k$  where  differences in  cosmic
evolution  have  imprinted a  different  signature  on  the growth  of
structure  on small  scales,  perhaps reflecting  the  effect of  dark
energy on conditions when structure formed at high redshift.

Those models that  show a greater non-linear growth  are those that in
the early  universe had a  greater contribution of the  matter density
relative to  the dark energy  density; these correspond to  the models
with  today $w_0>w_{\rm  eff}$ and  in  the recent  universe had  {\it
lower\/}  matter  density  relative  to  dark  energy  density.   This
suggests  that  non-linear  growth  is more  sensitive  to  conditions
(including the effects of dark  energy) in the early, matter dominated
universe than  it is  to conditions in  the later, accelerated  era of
dark energy domination.

Carrying this  forward, one conjecture  is that the  ``transition'' in
the  behavior of  the $z=0$  power spectrum  at $k=1\,\hovm$  might be
related to  early, rather than  $z=0$, non-linear effects.   While the
non-linear  scale at $z=0$  should be  near $k\lesssim0.2$,  the power
spectrum remains  well matched  here, possibly because  the non-linear
growth was  already matched  as a result  of the model  mapping, i.e.\
``pinned  down''  by  the  agreement  at $a=0.24$.   So  the  greatest
difference in  non-linear growth, arising from times  earlier than the
$a=0.24$  matching, might  appear at  the non-linear  scale associated
with $a<0.24$, or $k\gtrsim1\,\hovm$, rather than the $z=0$ non-linear
scale.  In any case, the  accurate approximation of the power spectrum
utilizing the  matching prescription indicates  that reasonably simple
physics lies behind even the non-linear mass power spectrum.

\section{Conclusion} \label{Conclusion}

The mass  power spectrum lies  at the foundation of  many cosmological
observables, such  as the weak  lensing shear statistics  of galaxies,
the  large scale structure  clustering distribution  (including baryon
acoustic  oscillations), and  cluster abundances.   To utilize  any of
these  cosmological measurements  as next  generation probes  of large
scale structure, cosmology, or dark energy requires clear and accurate
understanding  of the  mass power  spectrum over  the range  of models
under   consideration,   e.g.\   dynamical   dark  energy   not   just
$\Lambda$CDM.

The       main      results       of      this       article      (see
Figs.~\ref{wm1power},\ref{spower0} and \ref{spower3}) demonstrate that
non-linear  mass power spectra  of dynamical  dark energy  models with
smooth  equation  of state  evolution  can  be  determined to  percent
accuracy by calculating  the power for the constant  equation of state
cosmology that  gives a matching  distance to the CMB  last scattering
surface.   By varying  other parameters  as well,  such as  the matter
density $\om$  and $h$  keeping $\om h^2$  constant (see LW),  one can
envision mapping  a wide  variety of dark  energy models  to $\lam$CDM
models,  resulting in significant  gains in  computational efficiency.
This can also alleviate concerns regarding phantom crossing.

Finding the  distance matched models as  described in this  paper is a
trivial task  numerically requiring the integration  of a differential
equation and a one dimensional parameter search. This simple procedure
however provides  a mapping that is  accurate to a  percent. Of course
the accuracy of the  resultant power spectrum estimation is ultimately
only as good  as the accuracy in the model being  mapped to.  For that
model we  can then  utilise fitting formulas,  such as Halofit,  for a
rough,   $\sim10\%$  accuracy   or  more   generally   perform  N-body
simulations  with  the  desired  parameters.   However,  the  distance
matching scheme in this case allows a much reduced grid of simulations
to be carried out while still maintaining a high degree of accuracy.

The  physics behind  the  matching prescription  involves  a chain  of
consequences from the crossover in the behavior of one variable (e.g.\
equation  of state)  to the  convergence in  another, then  leading to
matching in  the large scale  structure growth.  Further  physics, not
yet  fully elucidated,  points to  the full  nonlinear  power spectrum
being dependent not  on the linear growth, but  the linear growth {\it
history\/},  where  the conditions  (e.g.\  matter  density or  growth
factor) at one epoch directly  manifest in the nonlinear behavior at a
later epoch.

Future work should  pursue this further, as well  as investigating the
prospects for discerning the  signatures of more complicated equations
of state  or perturbations.  While  the prescription here  for mapping
the  mass power spectrum  to percent  accuracy between  cosmologies is
useful  in  itself and  for  computational  gains,  the most  exciting
prospects  are  for  improved  analytic fitting  formulas  and  deeper
physical understanding.

\section*{Acknowledgments}

MF acknowledges  the support  of a Science  Faculty UPA,  thanks Chris
Power and Jeremy Bailin for  helpful discussions, advice and pieces of
code, and thanks LBNL and SNAP for hospitality and support during much
of the  writing of this article.   We thank Martin  White for pointing
out the nice  method of getting high resolution  FFT's without needing
large arrays  and for  other useful discussions.   This work  has been
supported by  the Australian Research  Council under grant  DP 0665574
and  in part  by the  Director, Office  of Science,  US  Department of
Energy  under  grant  DE-AC02-05CH11231.

\bsp

\label{lastpage}

\end{document}